\documentclass[%
reprint,
 showpacs,%preprintnumbers,
amssymb,
 aps,
pra,
 longbibliography
]{revtex4-1}

\usepackage{amsmath,dsfont}
\usepackage{mathbbol}
\usepackage{bbold}
\usepackage{mathrsfs}
\usepackage{graphicx}% Include figure files
\usepackage{dcolumn}% Align table columns on decimal point
\usepackage{bm}% bold math
\usepackage[usenames]{color}
\usepackage[normalem]{ulem}
\usepackage{hyperref}
\usepackage{comment}

\usepackage[per-mode=symbol]{siunitx}   
\usepackage[version=3]{mhchem}
 
\DeclareSIUnit\intensity{\watt\per\centi\meter\squared}
\DeclareSIUnit\fieldstrength{\volt\per\centi\meter}
\DeclareSIUnit\kfieldstrength{k\volt\per\centi\meter}
\DeclareSIUnit\energy{cm^{-1}}
\DeclareSIUnit\ev{eV}

\newcommand{\ie}{i.\,e.}%

\newcommand{\doubletseven}{\ensuremath{{}^2\text{S}^\text{e}}}
\newcommand{\doubletpodd}{\ensuremath{{}^2\text{P}^\text{o}}}

\newcommand{\doubletpe}{\ensuremath{{}^2\text{P}^\text{e}}}
\newcommand{\doubletdodd}{\ensuremath{{}^2\text{D}^\text{o}}}

\newcommand{\doubletd}{\ensuremath{{}^2\text{D}}}
\newcommand{\doubletde}{\ensuremath{{}^2\text{D}^\text{e}}}

\newcommand{\tripletpodd}{\ensuremath{{}^3\text{P}^\text{o}}}

\newcommand{\tripletpe}{\ensuremath{{}^3\text{P}^\text{e}}}

\newcommand{\qpe}{\ensuremath{{}^4\text{P}^\text{e}}}
\newcommand{\qsodd}{\ensuremath{{}^4\text{S}^\text{o}}}

\newcommand{\ket}[1]{\left|#1\right\rangle}

\newcommand{\mtot}{\ensuremath{M_\text{tot}}}

\newcommand{\ucm}{\affiliation{Departamento de Qu\'imica F\'isica, Universidad Complutense de Madrid, 28040 Madrid, Spain}}%
\newcommand{\aarphys}{\affiliation{Department of Physics and Astronomy, Ny Munkegade 120, Aarhus University, 8000  Aarhus C, Denmark}}%

\begin{document}

\title{Carrier-envelope phase effects in one- and two-photon directional photoionization of non-isotropic atomic states}

\author{Juan J.\ Omiste}\email{jomiste@ucm.es}\ucm
\author{Lars Bojer Madsen}\aarphys

\date{\today}
\begin{abstract}
  We study the impact of two-color ($\omega$ and $2\omega$) co- and counter-rotating ultrashort attosecond laser pulses on non-isotropic atomic targets through the one- and two-photon interference pattern of the photoelectron spectrum. Specifically, we take the ground state of atomic carbon,~\ie, $(1s^22s^22p^2,\tripletpe)$ as a prototype. We observe and quantify the strong dependency on the relative carrier-envelope phase (CEP) of the two-color pulses and on the spatial orientation of the electronic target states. Notably, we observe that the photoelectron momentum distributions (PMDs) vary as a function of the CEP due to the interfering two-color one- and two-photon ionization paths. Besides, the PMD region corresponding to one-photon photoionization remains unaffected, with varying CEP, depending only on the ellipticity of the pulse, the central photon frequency and the magnetic quantum number of the initial state. Therefore, comparing the one-photon ionization electron ejection direction following absorption of a single ($2\omega$) photon with that of the two-photon ionization channel following absorption of two photon each with energy $\omega$ we may extrapolate information on the CEP difference between the two pulses.
\end{abstract}

\maketitle

\section{Introduction}
\label{sec:introduction}

The study of photoelectron momentum distributions (PMDs) generated by ultrashort laser pulses has become a powerful tool to probe fundamental interactions in atomic and molecular systems. Advances in laser technology have enabled precise control over pulse characteristics, such as carrier-envelope phase (CEP), polarization, and spectral composition, paving the way to investigate the intimate nature of quantum phenomena. These advances have been particularly impactful in studying photoemission time delays, the design of attoclocks~\cite{Pfeiffer2013,Eicke2019}, and the study of non-isotropic atomic targets, where the spatial orientation of electronic states plays a crucial role~\cite{DeSilva2021,Omiste2021a}.

The interaction of ultrashort laser pulses with atoms and molecules has been extensively studied for both isotropic and oriented systems. In particular, bichromatic laser fields, especially those involving elliptically or circularly polarized light, have revealed rich dynamics, including circular dichroism~\cite{Davydiak2025} and vortex structures~\cite{NgokoDjiokap2015,Geng2020} in PMDs. Even relativistic descriptions of such interactions, for example, helium interacting with elliptically polarized bichromatic lasers, have shown how the interplay of photon energy, polarization, and CEP leads to distinct features in ionization pathways~\cite{Hofbrucker2021}. Similarly, experimental investigations, such as those involving barium in excited states~\cite{Yamazaki2007} and lithium~\cite{Acharya2021}, have provided critical insights into the role of the initial state and laser polarization in shaping the photoelectron spectrum.

Furthermore, the handedness of circularly polarized light implies circular dichroism, a different behavior of the enantiomers of chiral species, allowing the exploration in atomic and molecular systems~\cite{DeSilva2021a,Wagner2024}. For instance, resonance-enhanced few-photon ionization studies have demonstrated the sensitivity of PMDs to the CEP and polarization state~\cite{DeSilva2021a}. In molecules, such as H${}_2^+$, bicircular laser fields have been shown to induce complex vortex structures and control photoelectron emission directions~\cite{Beaulieu2017,Beaulieu2018,Yuan2016,Yuan2017,Liu2023,NgokoDjiokap2015,Geng2020}.

Nowadays, the unprecedent sensitivity of the interferometry at the attosecond scale allows to track the impact of the electronic correlations on these processes. Hence, the use of theoretical and numerical methods accounting for it accurately are mandatory, even for non-isotropic atomic species~\cite{Omiste2021a}. These methods include multireference configuration interaction, multiconfigurational Hartree-Fock~\cite{Hochstuhl2010} or other time-dependent \textit{ab initio} methods~\cite{Sato2013,Miyagi2013,Omiste2017_be}. In this context, we study the ground state of atomic carbon, given by the non-isotropic open-shell triplet state $(1s^22s^2p^2,\tripletpe)$. The degeneracy of this state will allow to compute the PMDs for different total magnetic quantum number or orientation of the electronic cloud of the target state. Note that recent studies have demonstrated the capability to control the population of atomic states according to magnetic quantum numbers using external magnetic fields~\cite{Brouard2014, Gordon2018}, laying the groundwork for experimentally investigating these systems under laser-driven ionization conditions. 

We focus on the interaction of two-color co- and counter-rotating attosecond laser pulses with non-isotropic atomic carbon. The two colors have frequencies $\omega_1=\omega$ and $\omega_2=2\omega$, such that one- and two-photon ionization pathways may interfere by absorption of either a single photon of energy $\omega_2$ or two photons each of energy $\omega_1$. We make use of the time-dependent restricted-active-space self-consistent-field (TD-RASSCF) methodology~\cite{Miyagi2013,Miyagi2014,Miyagi2014b,Omiste2017_be}, which correctly captures the electronic correlation while using a reduced number of configurations~\cite{Omiste2018_neon}. We study the photoelectron interference patterns arising from one- and two-photon ionization pathways. By examining the dependency of PMDs on the CEP and the spatial orientation of the initial electronic state, we aim to extract the role of these factors in determining the ionization dynamics.

The paper is organized as follows: First, in Sec.~\ref{sec:theory_and_methods} we briefly introduce the TD-RASSCF theory and methodology applied for the computations.  Next, in Sec.~\ref{sec:discussion_and_results} we describe the system under study and present the main results. Finally, in Sec.~\ref{sec:conclusions_and_outlook} we give the main conclusions and an outlook for future investigations. 
\section{Theory and methods}
\label{sec:theory_and_methods}
We employ the TD-RASSCF methodology~\cite{Miyagi2013,Miyagi2014,Miyagi2014b} to describe both the initial states~\cite{Omiste2021a} and the dynamics following ionization~\cite{Madsen2018,Lode2020}. Here, we briefly describe the TD-RASSCF methodology, devoted to solve the time-dependent Schr\"odinger equation for an $N$-body fermionic system; for formulation for bosons and extension multispecies, see Refs.~\cite{Leveque2016a,Leveque2018,Leveque2019}. We take as an ansazt to the electronic wave function
\begin{equation}
\label{eq:wf_ansazt}
\ket{\Psi(t)}=\sum\limits_{\mathcal{I}}C_\mathcal{I}(t)\ket{\Phi_\mathcal{I}(t)},
\end{equation} 
where the summation runs over configurations $\mathcal{I}$ represented by Slater determinants, $\ket{\Phi_\mathcal{I}(t)}$. Each $\ket{\Phi_\mathcal{I}(t)}$ is formed by single-electron time-dependent orbitals $\phi_{\mathcal{I},j}(t)$, and $C_\mathcal{I}(t)$ is a time-dependent amplitude. The corresponding equations of motion can be found  in the references given above. The TD-RASSCF method considers that the active orbital space is divided in three parts: frozen orbitals in $\mathcal{P}_0$, active orbitals in $\mathcal{P}_1$ and excited orbitals in $\mathcal{P}_2$. In our implementation, the single-electron orbitals are represented by finite element discrete-variable-representation (FE-DVR) functions~\cite{McCurdy1991,Omiste2017_be} times a spherical harmonic function. The time-dependency of the orbitals allows to work with a small set of configurations compared to methodology based on configuration interaction with time-independent orbitals~\cite{Omiste2019}. In this work, we consider 5 orbitals belonging to the $\mathcal{P}_1$ partition of the active space, hence, all the combinations of orbitals are allowed, which is equivalent to the multiconfigurational time-dependent Hartree-Fock (MCTDHF) approach~\cite{Hochstuhl2014,Madsen2018,Lode2020}.

We obtain the initial wave function by performing a propagation in imaginary time of a trial function~\cite{Miyagi2013,Omiste2017_be} which fullfils the symmetry restrictions of the ground state~\cite{Omiste2018_neon, Omiste2021a}. Since we work with atomic carbon we set the trial function so that it includes the orbitals $1s$, $2s$, $2p_{\pm 1}$ and $2p_0$ and only the amplitudes of the configurations with the required symmetry are nonzero. Specifically, we impose that the total magnetic quantum number is fixed and the parity is even, as required for the \tripletpe~ground state of carbon~\cite{Omiste2018_neon}.

Let us now briefly describe the general features of the photoionization process, following from one-photon absorption by C$(\tripletpe)$, leading to C${}^+$ left in the state $\mathcal{T}$. In this case, the ionization reaction is described by
\begin{equation}
    \label{eq:ionization_reaction_general}
\text{C}(1s^22s^22p^2[\tripletpe])[\mtot]+\gamma\rightarrow \left[\text{C}^+[\mathcal{T}][\mtot^\prime]+e^-(\ell m)\right]{}^3L^\text{o},  
\end{equation}
where $\mtot=0,\,\pm 1$ corresponds to the total magnetic quantum number of the ground state of the carbon atom. $L$ stands for the total angular momentum of the C$^+$ and the outgoing electron, and $\mtot^\prime$ is the total magnetic quantum number of $\text{C}^+$ in the state $\mathcal{T}$ and the pair $(\ell m)$ describes the orbital angular momentum and magnetic quantum number of the outgoing electron, respectively. According to the electric dipole selection rules, $L=0,1,2$ since $\Delta L=\pm 1\, (L=0\leftrightarrow L^\prime=0~\text{forbidden})$ and  $\mtot=\mtot^\prime+m$ for linearly polarized along the $Z$ axis and $\mtot=\mtot^\prime+m\pm 1$ for circularly polarized light on the $XY$-plane.  Note that the term on the rhs of Eq.~\eqref{eq:ionization_reaction_general} includes information about the parity which has changed from even (e) to odd (o).

\begin{figure}[b]
	\includegraphics[width=.9\linewidth]{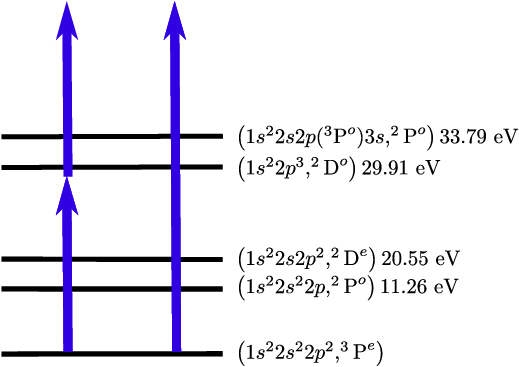}
	\caption{\label{fig:fig1} Energy levels of the states involved in the photoionization of the ground state of C(\tripletpe) along with one- and two-photon absorption leading to interference. The central photon energies used correspond to $50$~eV and $25$~eV, respectively, to have equal duration. See also Table~\ref{tab:ionization_energies}.}
\end{figure}

We also analyze the case of two-photon absorption, mandatory to understand the interference of photoelectrons in the two-color co- and counter-rotating laser pulses. For that case, we assume that the predominant process led by the second photon consists of transferring energy and angular momentum to the electrons ejected from the carbon ground state. This process is described in the following reaction
\begin{eqnarray}
\nonumber
 \text{C}([\tripletpe])[\mtot]+\gamma+\gamma^\prime&\rightarrow& %\text{C}^+[\mathcal{T}][\mtot^\prime]+e^-(\ell m)+\gamma^\prime\\
%&\rightarrow&
 \left[\text{C}^+[\mathcal{T}][\mtot^\prime]+e^-(\ell^\prime m^\prime)\right]{}^3{L^\prime}^\text{e},
\end{eqnarray}
with $L^\prime=0,1,2, 3$ and $\mtot^\prime+m^\prime-\mtot=0,\pm 1,\pm 2$. In this work we do not consider double photoionization, which occurs with negligible probability compared to the single-ionization. Note that the energy of a single photon is above the first ionization threshold [See Fig.~\ref{fig:fig1}].

We model the laser interaction in the velocity gauge. The attosecond XUV pulse is described by a two-color combination of vector potentials $\vec{A}=\vec{A}_1+\vec{A}_2$, with components of the form
\begin{equation}
    \label{eq:vector_potential}
    \vec{A}_j=\frac{A_{0,j}}{\sqrt{1+\varepsilon_j^2}}\cos^2[\omega_j t/(2 n_{p,j})]\left(
    \begin{array}{c}
         \varepsilon_j\cos(\omega_j t+\varphi_j) \\
         0 \\
         \sin(\omega_j t+\varphi_j)\\
    \end{array}
    \right),
\end{equation}
where $A_{0,j}$ corresponds to a maximum intensity of  $\SI{e14}{\intensity}$, $\varepsilon_j$ is the ellipticity, $n_{p,j}$ the number of cycles and $\varphi_j$ denotes the carrier-envelope phase (CEP) for component $j$. The components $\vec{A}_1$ and $\vec{A}_2$ are set so that they are turned on at time $t=-\frac{T}{2}$ and turned off at $t=\frac{T}{2}$ with $T=\frac{2\pi n_{p,1/2}}{\omega_{1/2}}$ and $\omega_2=2\omega_1$. 

The analysis of the photoelectron momentum distributions (PMDs) is performed by projecting the time-dependent wave function in the outer region on the Coulomb functions~\cite{Madsen2007,Yu2017,Omiste2018_neon}.

\section{Discussion and Results}
\label{sec:discussion_and_results}
We explore the laser-induced ultrafast ionization of the carbon ground state by means of co- and counter-rotating pulses. We assume that the polarization plane is contained in the $XZ$ plane [see Eq.~\eqref{eq:vector_potential}]. Since the \tripletpe~ states are not invariant under rotations around the $Y$ axis,~\ie, the propagation direction of the laser pulse, the electric field of the laser sees the anisotropy of the ground state as an imbalanced charge distribution. Let us briefly describe this feature.%, as shown in the sketch of the wave functions for $\mtot=1$ and $0$ in Figs.~\ref{fig:c_2p} and~\ref{fig:c_2pabs1}. There, we observe that the charge distribution for $\mtot=\pm 1$ is located along the $Z$ axis of the lab-fixed frame and the $XY$ plane (Fig.~\ref{fig:c_2pabs1}), whereas it is mainly located around the $XY$ plane for the $\mtot=0$ state. 

The main configuration of the ground state of C is $1s^22s^22p^2$. Taking into account that the $1s$ and $2s$ subshells are full, we may in a simple analysis express the $\tripletpe\left[\mtot\right]$ state as a function of $2p$-electron wave functions as
\begin{align}
\label{eq:gs_configuration_mm1}
  \tripletpe\left[-1\right] \rightarrow& \cfrac{1}{\sqrt{2}}\left(\phi_{p_{0}}\phi_{p_{-1}}-\phi_{p_{-1}}\phi_{p_{0}}\right){\chi}^{1}_{M_S},\\
\label{eq:gs_configuration_m0}
  \tripletpe\left[0\right] \rightarrow &  \cfrac{1}{\sqrt{2}}\left(\phi_{p_{1}}\phi_{p_{-1}}-\phi_{p_{-1}}\phi_{p_{1}}\right){\chi}^{1}_{M_S},\\
\label{eq:gs_configuration_m1}
  \tripletpe\left[1\right] \rightarrow &  \cfrac{1}{\sqrt{2}}\left(\phi_{p_{0}}\phi_{p_{1}}-\phi_{p_{1}}\phi_{p_{0}}\right){\chi}^{1}_{M_S},
\end{align}
where ${\chi}^{1}_{M_S}$ denotes the triplet spin state characterized by the projection of the spin along the $Z$-axis, $M_S$, and where $\phi_{p_j}(j=0,\pm 1)$ denotes a spatial orbital. In this representation, the atomic terms within the triplet mix under rotations around any axis other than $Z$ of the laboratory fixed frame (LFF)~\cite{Zare1988}. With these insights, let us now describe the shape of the electronic cloud for each state in the \tripletpe ~term. First, note that for $\mtot=0$ the electrons are located in the $XY$-plane since the wave function is formed by $p_{\pm 1}$ orbitals, see Eq.~\eqref{eq:gs_configuration_m0}. Hence, we expect that this state is less affected by any $Z$-component of the laser pulse. On the other hand, one electron in the $\mtot=\pm 1$ case is located close to the $XY$-plane, whereas the other electron is in the $p_0$ orbital,~\ie, around the $Z$ axis, see Eqs.~\eqref{eq:gs_configuration_mm1} and~\eqref{eq:gs_configuration_m1}. These characteristics of the initial state determine the shape of the PMDs, as we discuss in detail in Sec.~\ref{sec:pmds_structure}.

Eqs.~\eqref{eq:gs_configuration_mm1}-\eqref{eq:gs_configuration_m1} highlight that the atomic carbon ground state is open-shell, hence more than one Slater determinant is required to appropiately describe it. Therefore, a post Hartree-Fock method, such as the TD-RASSCF method~\cite{Miyagi2013}, needs to be applied. Our implementation of TD-RASSCF is restricted to an equal number of spin up and down electrons, therefore, this work focus on the $M_S=0$ component of $\tripletpe\left[\mtot\right]$. However, our conclusions are equivalent for $M_S=\pm 1$, since we do not consider the spin-orbit interaction or the magnetic component of the laser field.
%For the sake of clarity, we include a sketch of the wave functions in ~\autoref{fig:c_2pabs1} and~\autoref{fig:c_2p}, where we represent the density of $p_0$ ($p_z$) and $p_{\pm 1}$. 

\subsection{Structured PMD in circularly polarized pulses}
\label{sec:pmds_structure}

We now analyze the PMDs following the photoionization of the directional ground states of carbon by means of circularly polarized, co- and counter-rotating XUV pulses. We focus on the effects provoked by one- and two-photon ionization coherence as well as influence by the spatial orientation of the electronic cloud.

\subsubsection{Circularly polarized laser pulse}
\label{sec:circularly_polarized_laser_pulse}
\begin{figure*}
	\includegraphics[width=.7\linewidth]{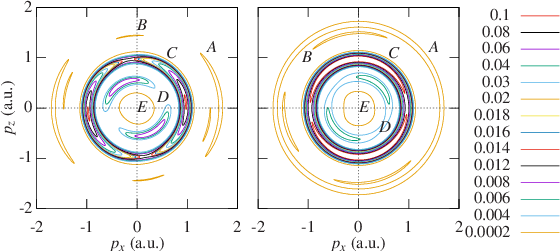}	
		\caption{\label{fig:fig2} Photoelectron Momentum Distribution of the ground state of C(\tripletpe)[$M_\text{tot}=0$] (left panel) and C(\tripletpe)[$M_\text{tot}=1$] ($\times 1.5$, right panel) after interacting with a one-color circularly polarized XUV pulse contained in the $XZ$-plane with central photon energy $25$~eV, 10 cycles duration and an intensity of $\SI{e14}{\intensity}$. The ring-shaped features corresponding to the photoionization channels are labeled as $A)$: two photon and $\text{C}^+[(1s^22s^22p)\doubletpodd]$ $B)$: two photon and $\text{C}^+[(1s^22s2p^2)\doubletde]$;  $C)$: one photon and $\text{C}^+[(1s^22s^22p)\doubletpodd]$; $D)$: one photon and $\text{C}^+[(1s^22s2p^2)\doubletde]$; $E)$: low energy photoelectrons from shake-up states (details are given in the main text). On the right we show the value of each level of the contour plots. Values of the signal in the PMD in the right panel is multiplied by 1.5 for the sake of clarity. }	
\end{figure*}

First, we consider the case of a one-color circularly polarized pulse of 10 cycles, $\omega=\SI{25}{\eV}$ and $I=\SI{e14}{\intensity}$. The PMDs are shown in Fig.~\ref{fig:fig2} for $\mtot=0$ and $1$. For $\mtot=0$ we clearly distinguish 5 rings, labeled as $A$, $B$, $C$, $D$ and $E$ in Fig.~\ref{fig:fig2}. As we have briefly discussed in Sec.~\ref{sec:theory_and_methods}, the rings appearing in the PMD correspond to ejected electrons formed during the photoionization process. In order to understand these features, such as shape and kinetic energy, and also identify the photoionization channels, we collect the main ionic species and the corresponding ionization thresholds in Table~\ref{tab:ionization_energies}.

\begin{table}%[htbp]
    \caption{\label{tab:ionization_energies} Low-lying ionization thresholds for C~\cite{NIST_ASD_2019}. The states are labeled using their main configurations and terms. Ionization threshold are given in units of eV.}
    \begin{ruledtabular} 
            \begin{tabular}{ll|ll}
                 $\text{C}^+$ state & Ionization & $\text{C}^+$ state & Ionization \\
				 &  threshold &  &  threshold \\
                \hline
                $(1s^22s^22p,\doubletpodd)$ & 11.26 & $(1s^22s^23s,\doubletseven)$ & 25.71 \\ % Replace with your data
                $(1s^22s2p^2,\qpe)$ & 16.59 & $(1s^22s^23p,\doubletpodd)$ & 27.59 \\ % Replace with your data
                $(1s^22s2p^2,\doubletde)$ & 20.55 & $(1s^22p^3,\qsodd)$ & 28.86 \\ % Replace with your data
                $(1s^22s2p^2,\doubletseven)$ & 23.22 & $(1s^22p^23d,\doubletde)$ & 29.31 \\
                $(1s^22s2p^2,\doubletpe)$ & 24.97 & $(1s^22p^3,\doubletdodd)$ & 29.91 \\                
 & & $(1s^22p^3,\doubletpodd)$& 32.67 \\           
 & & $(1s^22s2p(\tripletpodd)3s,\doubletpodd)$& 33.79 
            \end{tabular}
    \end{ruledtabular}
\end{table}

Then, we identify rings $C$ and $A$ with the photoionization through the channel $\text{C}^+[(1s^22s^22p)\doubletpodd]$ after absorbing one and two photons, respectively.

Specifically, the absorption of one photon leads to the ejection of $s$ and $d$ electrons, whereas the absorption of the second photon change the angular momentum to $\ell=1$ and $3$. As described at the beginning of Sec.~\ref{sec:discussion_and_results}, the electrons in the $2p$ shell for $\mtot=0$ are located around the $XY$-plane, hence, we expect the photoelectron in ring $C$ to be located close to the $X$ axis. This is the case for $\mtot=0$, as seen in Fig.~\ref{fig:fig2}, although it is slightly tilted due to the circular polarization of the laser. On the contrary, we observe that ring $C$ lies along the $Z$-axis for $\mtot=1$, due to the location of the $p_0$ electron close to the $Z$ axis of the LFF.

Next, ring $D$ corresponds to the photoionization through the channel $\text{C}^+[(1s^22s2p^2)\doubletde]$ leading to the photoemission of a $p$-wave electron. We observe that the photoemission is tilted in opposite directions for the cases $\mtot=0$ and $1$. This may be puzzling, since in both cases an electron from the $2s$ shell is removed. However, the electron density in the $2p$ shell distorts differently the final outcome. Let us also remark that ring $B$ corresponds to electrons from the same ionization channel after absorbing a second photon, leading to $\ell=0$ and $2$. 

 Finally, ring $E$ corresponds to low energy electrons coming from shake-up states, including $\text{C}^+[(1s^22p^3)\qsodd]$, $\text{C}^+[(1s^22p^3)\doubletpodd]$, $\text{C}^+[(1s^22p^3)\doubletdodd]$ and $\text{C}^+([1s^22s2p(\tripletpodd)3s]\doubletpodd)$. Note that their ionization threshold (above \SI{29.91}{\ev}, see~\autoref{tab:ionization_energies} and Fig.~\ref{fig:fig1}) is higher than the central photon frequency of the laser, hence the photoionization is feasible only due to the laser bandwidth.

\subsubsection{Two-color corotating laser pulse}
\label{sec:corotating_laser_pulse}
\begin{figure}[b]
	\includegraphics[width=.95\linewidth]{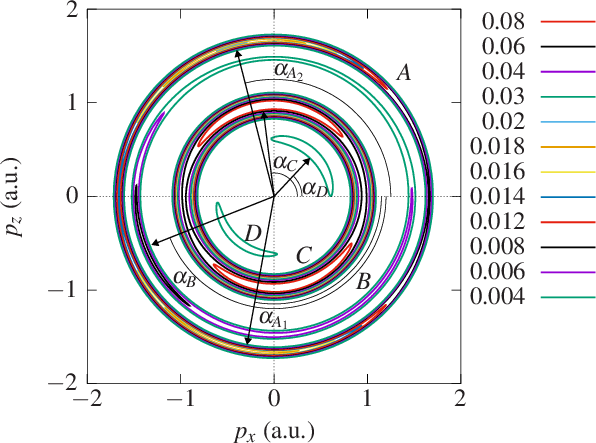}
	\caption{\label{fig:fig3} Photoelectron Momentum Distribution of the ground state of C(\tripletpe)[$M_\text{tot}=1$] after interacting with a two-color co-rotating pulse contained in the $XZ$-plane for $\varphi=0$, for the central photon energies $50$~eV and $25$~eV and the number of cycles $20$ and $10$, respectively. The rings are labeled as in Fig.~\ref{fig:fig2}. The local maxima within each ring are highlighted by arrows and the angle formed by them with the $X$ axis is also shown for clarity. %In the case there are more than one maxima per ring, we enumerate them arbitrarily, corresponding ring $1$ to solid, $2$ to dashed and $3$ to dash-dotted arrow. 
	The intensity of both pulses is $\SI{e14}{\intensity}$.  On the right of the plot we show the value of each contour, as in Fig.~\ref{fig:fig2}. }
\end{figure}

\begin{figure*}
	\includegraphics[width=.8\linewidth]{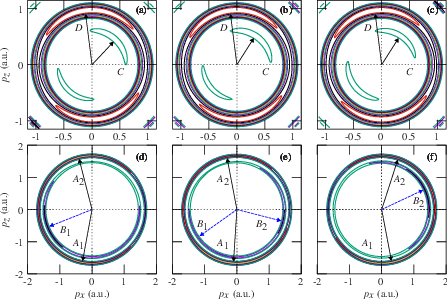}
	\caption{\label{fig:fig4} Photoelectron Momentum Distribution of the ground state of C(\tripletpe) with $M_\text{tot}=1$ after interacting with a two-color co-rotating pulse contained in the $XZ$-plane. The intensity of the pulses is set to \SI{d14}{\intensity} and the central photon energies to $50$~eV and $25$~eV, being the number of cycles $20$ and $10$, respectively. The CEP of the \SI{25}{\ev} pulse, $\varphi$, is set to $\varphi=0$ [(a) and (d)], $\pi/5$ [(b) and (e)] and $2\pi/5$ [(c) and (f)]. The photoionization channels $C$ and $D$ are shown in panels (a)-(c), and channels $A$ and $B$ in panels (d)-(f), as described in Fig.~\ref{fig:fig2} and~\ref{fig:fig3}. The local maxima within rings $A$ (solid black) and $B$ (dashed blued) are pointed within the rings. The contour lines correspond to the same values as in Fig.~\ref{fig:fig3}. 
}
\end{figure*}
Now, we address the case of superimposing two circularly polarized laser pulses, specifically, $\omega=\SI{25}{}$ and $\SI{50}{\ev}$ with 10 and 20 cycles, respectively, such that the pulses have equal duration.

First, we describe the labeling within the rings in the PMDs. The labeling is sketched for the corotating case with $\varphi=0$ in Fig.~\ref{fig:fig3}. We identify the peaks using the ionization channel and the increasing order of angle formed with the $X$ axis, $\alpha\in(-\pi,\pi]$. For instance, there are two peaks within ring $A$, pointed with arrows and labeled by the angle formed with the $X$ axis. Peak $A_1$ forms an angle $\alpha_{A_1}\approx -1.75$~rad whereas in the upper half of the PMDs we observe $A_2$, with $\alpha_{A_2}\approx 1.82$~rad. Let us remark that we only consider one peak within rings $C$ and $D$, since the other peaks are obtained by a rotation by $\pi$. The ring $E$ is not shown for the sake of clarity.

Now, we describe in detail the dependence on the CEP of the PMD in the corotating case by setting $\varepsilon=1$ in Eq.~\eqref{eq:vector_potential} for each circularly polarized component. The CEP, $\varphi$, [see Eq.~\eqref{eq:vector_potential}] is set to $0$ for the \SI{50}{\ev} component, and varied for the \SI{25}{\ev}. The PMDs for $\mtot=1$ in Fig.~\ref{fig:fig4} show the 4 well-defined rings, corresponding to the photoionization channels described in Sec.~\ref{sec:circularly_polarized_laser_pulse}. Specifically, in Fig.~\ref{fig:fig4}(a)-(c) we highlight the PMDs corresponding to rings $C$ and $D$ for $\varphi=0,\pi/5$ and $2\pi/5$, respectively. Besides, in Fig.~\ref{fig:fig4}(d)-(f) we plot the rings $A$ and $B$ for the same values of~$\varphi$. Note that ring $E$ is not shown for the sake of clarity. The outermost ring,~\ie, ring $A$, corresponds to the interference of $s$ and $d$-wave electrons from one-photon ionization through channel $\text{C}^+[(1s^22s^22p), \doubletpodd]$ and the $p$ and $f$-wave electron from two-photon ionization by means of \SI{25}{\ev}, as depicted in Fig.~\ref{fig:fig1}. The narrow ring inside this one (ring $B$) corresponds to the interaction of $p$-wave electrons from one-photon ionization through the channel $\text{C}^+[(1s^22s2p^2), \doubletde]$ and the two-photon ionization of $s$ and $d$-wave electrons. On the other hand, the absorption of a single \SI{25}{\ev} photon leads to the two inner-most rings (rings $C$ and $D$), as described in Sec.~\ref{sec:circularly_polarized_laser_pulse}. Rings $C$ and $D$ correspond to one-photon ionization, therefore, they are weakly affected by CEP effects, as it is well-known in lowest-order perturbation theory. Therefore, the angle formed by the peak of rings $C$ and $D$ with the $X$ axis, denoted as $\alpha_C$ and $\alpha_D$, are also independent of $\varphi$. Specifically, $\alpha_C=1.70$ for $\varphi=0,\pi/10,\pi/5,3\pi/10,2\pi/5$ and $\pi/2$. On the contrary, $\alpha_D$ slightly varies, being $0.80$ and $0.88$ for $\varphi=0$ and $\pi/10$, remaining $\alpha_D=1.00$ for the rest of the values analyzed.

%\begin{figure}[b]
%\includegraphics[width=.95\linewidth]{./FIGURES/ring_a_b_corrotating_angle}
%\caption{\label{fig:ring_a_b_corrotating_angle} Inclination angle of the maximum ejection, $\alpha_A$ and $\alpha_{B_j}$ ($j$th maximum) for the carbon ground state with $\mtot=1$ after interacting with the corotating pulse described in Sec.~\ref{sec:corotating_laser_pulse}.}
%\end{figure}

\begin{figure*}
	\includegraphics[width=.8\linewidth]{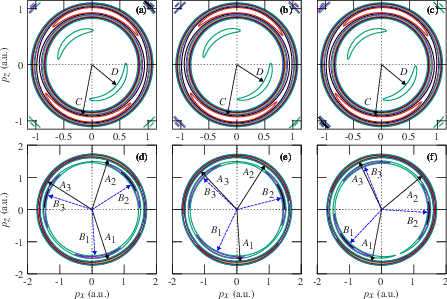}	
\caption{\label{fig:fig5}Photoelectron Momentum Distribution of the ground state of C(\tripletpe) with $M_\text{tot}=1$ after interacting with a two-color counter-rotating pulse contained in the $XZ$-plane. The intensity of the pulses is set to \SI{d14}{\intensity} and the central photon energies to $50$~eV and $25$~eV, being the number of cycles $20$ and $10$, respectively. The CEP of the \SI{25}{\ev} pulse, $\varphi$, is set to $\varphi=0$ [(a) and (d)], $\pi/5$ [(b) and (e)] and $2\pi/5$ [(c) and (f)]. The photoionization channels $C$ and $D$ are shown in panels (a)-(c), and channels $A$ and $B$ in panels (d)-(f). The local maxima within rings $A$ (solid black) and $B$ (dashed blue) are pointed within the rings.  
The contour lines correspond to the same values as in Fig.~\ref{fig:fig3}. 
}
\end{figure*}

Next, we describe rings $A$ and $B$. First, we consider ring $A$, corresponding to the remaining ion $\text{C}^+[(1s^22s^22p),\doubletpodd]$. The interference terms correspond to photoelectrons with $\ell=0$ and $2$ from the \SI{50}{\ev} photoionization and $\ell=1$ and $\ell=3$, coming from the two-photon absorption. The distribution within ring $A$ is characterized by two lobes, where the position of its maximum on the upper half of the PMD, $\alpha_{A_2}$, varies with the CEP. Ring $A$ presents a two-fold symmetry around $\pi/2$ ranging from approximately $1.26$ to $1.92$ for the chosen values of $\varphi$.

Now we turn to ring $B$, corresponding to one and two-photon ionization through the channel $\text{C}^+[(1s^22s2p^2),\doubletd]$. The ionization threshold is \SI{20.5}{\ev}, hence, the photoionization yield for the absorption of the \SI{25}{\ev} photon is high. The absorption of the second photon leads to photoelectrons of \SI{29.6}{\ev} described by $\ell=0$ and $2$. On the other hand, it interferes with the $p$ photoelectron ejected by the $\SI{50}{\ev}$ photon. We observe a dominant peak labeled as $B_1$. Its position varies smoothly for the CEP ranging from $0$ to $\pi/5$, and gets wider and blurred as we further increase $\varphi$. We clearly see another peak at $\varphi=\pi/5$ that we identify with $B_2$. Its position ranges from $\alpha_{B_2}\approx 0$ to $0.87$. We observe that this structure is very blurry, as well. This is consistent with previous works in the single active electron approximation~\cite{Yuan2016}.

\subsubsection{Two-color counter-rotating laser pulse}
\label{sec:counter-rotating_laser_pulse}
Next, we complement the description by including the counter-rotating case, shown in Fig.~\ref{fig:fig5}. We keep $\varepsilon=1$ for $\omega=\SI{50}{\ev}$ and change to ellipticity to $\varepsilon=-1$ for the component with $\omega=\SI{25}{\ev}$. Let us remark that the rings $C$ and $D$ are tilted in the opposite direction as observed in the co-rotating case [Fig.~\ref{fig:fig4}], since the \SI{25}{\eV} pulse has opposed ellipticity compared to the co-rotating case. As we dicussed before, the positions of the maxima within rings are almost invariant under the change of $\varphi$, as shown in both Fig.~\ref{fig:fig4} and~\ref{fig:fig5}. This fact allows to use the position of these maxima as an offset for the photoemission angle.

\begin{figure}
\includegraphics[width=.95\linewidth]{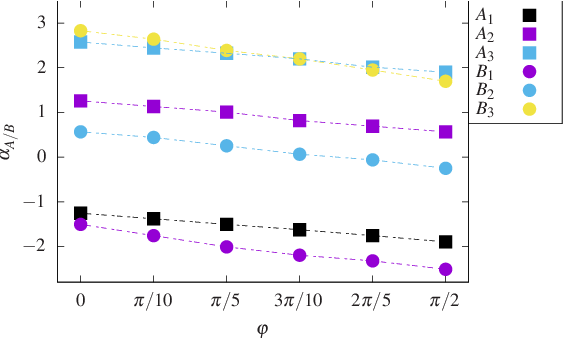}
\caption{\label{fig:fig6} Inclination angle of the maximum ejection, $\alpha_{A_j}$ and $\alpha_{B_j}$ ($j$th maximum) for the carbon ground state with $\mtot=1$ after interacting with the counter-rotating pulse described in Sec.~\ref{sec:corotating_laser_pulse}.}
\end{figure}

On the one hand, the ring $A$ is expected to present a four lobes structure due to the interference pattern of photoelectrons with $\ell$ from 0 to 3. However, this structure is not clearly observed, for instance, for $\varphi=\pi/5$ only 3 lobes can be distinguished. Therefore, it may not be considered a good candidate to account for the angle between photoelectrons from different channels. The values of the three local maxima of ring $A$ in Fig.~\ref{fig:fig5}, $\alpha_{A_{j=1,2,3}}$, are shown in Fig.~\ref{fig:fig6}. We observe approximately the same trend as a function of $\varphi$, keeping an almost constant angle between them.

On the other hand, ring $B$, in Fig.~\ref{fig:fig5}, shows the typical three lobes structure since the interference pattern is due to the photoionization of an electron in the $2s$ shell. First of all, note that the emission angles for $\varphi=0$ are $-1.51$, $0.57$ and $2.83$, being the angle between them $2.07$, $2.26$ and $1.95$, respectively. Let us remark that the differences do not correspond to exactly $2\pi/3$ due to the nonisotropic distortion of the $2p$ shell. These differences remain almost constant since the variation is similar for all $\alpha_{B_j}$ in Fig.~\ref{fig:fig6}. For instance, the different angle between the local maxima corresponding to photoionized electrons are $\varphi=3\pi/10$ are $2.26$, $2.14$ and $1.88$.

As in the corotating case, shown in Fig.~\ref{fig:fig4}, the relative orientation of the inner rings ($C$ and $D$) with respect to the outer rings ($A$ and $B$) varies as a function of $\varphi$. However, the ring shaped structures in the counter-rotating case are more stable,~\ie, the three-fold structure of rings $A$ and $B$ remains almost unaltered. As a consequence, the variation of $\varphi$ nicely correlates to the rotation of the PMDs substructures, as observed in Fig.~\ref{fig:fig6}. This feature allows to use the orientation of the rings associated to one-photon ionization as the balancing points for the orientation of the local maxima associated to the one- and two-photon interference terms. This implies accuracy in the measurements which may lead to a high sensitivity on the laser parameters.

\section{Conclusions and outlook}
\label{sec:conclusions_and_outlook}
In this work we have used the post Hartree-Fock, time-dependent many-electron theory, TD-RASSCF, to study in detail the interference pattern of one- and two-photon ionization mediated by two-color co- and counter-rotating laser pulses on a non-isotropic open-shell atomic target, specifically, the ground state of atomic carbon. We have described the photoelectron momentum distribution after photoionization, focusing on the carrier-envelope phase effects. On the one hand, the direction of the electron ejected after one-photon ionization remains almost unaltered, although it depends on the polarization of the XUV pulse and spatial orientation of the initial ground state. On the other hand, the interference of one- and two-photon ionized electrons is strongly affected by the carrier-envelope phase of the incident XUV pulse. We outline that by using the main one-photoionization direction as defining an offset angle, the rotation angle of the one- and two-photoionization interference pattern could be measured and related to the carrier-envelope phase.

Let us also remark that previous studies have restricted their investigation on non-isotropic polarization to one-electron molecules, such as H$_2^+$ or H$_3^{2+}$~\cite{Yuan2016,Yuan2017,Liu2023} or within the single-active-electron approximation. On the contrary, our formalism allows the extension to few electrons systems, such as anisotropic atomic species, for instance oxygen or boron or to few electron molecules, including H$_2$, Li$_2$ or LiH, allowing the study of electronic correlation using non-linearly polarized sources.

\begin{acknowledgments}
The numerical results presented in this work were obtained at the Centre for Scientific Computing, Aarhus. J.J.O. acknowledges the funding by the Madrid Government
(Comunidad de Madrid Spain) under the Multiannual Agreement with Universidad Complutense de Madrid in the line Research Incentive for Young PhDs, in the context of the V PRICIT (Regional Programme of Research and Technological Innovation) (Grant: PR27/21-010) and Projects PID2019-105458RB-I00, PID2021-122839NB-I00 and PID2022-138288NB-C33 (MICIN).
\end{acknowledgments}

%\clearpage
%\newpage

%\bibliography{time_dependent_many_e}

%

\end{document}